\documentclass[10pt,twocolumn,letterpaper]{article}

\usepackage[pagenumbers]{wacv} 

\usepackage{graphicx}
\usepackage{amsmath}
\usepackage{amssymb}
\usepackage{booktabs}

\usepackage[pagebackref,breaklinks,colorlinks]{hyperref}

\usepackage[capitalize]{cleveref}
\crefname{section}{Sec.}{Secs.}
\Crefname{section}{Section}{Sections}
\Crefname{table}{Table}{Tables}
\crefname{table}{Tab.}{Tabs.}

\def\wacvPaperID{2167} 
\def\confName{WACV}
\def\confYear{2025}

\begin{document}

\title{Motion Free B-frame Coding for Neural Video Compression}

\author{Van Thang Nguyen\\
FPT Software, AI Center\\
Hanoi, Vietnam\\
{\tt\small ThangNV108@fpt.com}
}
\maketitle

\begin{abstract}
Typical deep neural video compression networks usually
follow the hybrid approach of classical video coding that
contains two separate modules: motion coding and residual
coding. In addition, a symmetric auto-encoder is often used
as a normal architecture for both motion and residual coding. In this paper, we propose a novel approach that handles
the drawbacks of the two typical above-mentioned architectures, we call it kernel-based motion-free video coding. The
advantages of the motion-free approach are twofold: it improves the coding efficiency of the network and significantly
reduces computational complexity thanks to eliminating
motion estimation, motion compensation, and motion coding which are the most time-consuming engines. In addition, the kernel-based auto-encoder alleviates blur artifacts
that usually occur with the conventional symmetric autoencoder. Consequently, it improves the visual quality of the
reconstructed frames. Experimental results show the proposed framework outperforms the SOTA deep neural video
compression networks on the HEVC-class B dataset and is
competitive on the UVG and MCL-JCV datasets. In addition, it generates high-quality reconstructed frames in comparison with conventional motion coding-based symmetric
auto-encoder meanwhile its model size is much smaller than
that of the motion-based networks around three to four
times.
\end{abstract}

\section{Introduction}
\label{sec:intro}

Video coding has a long history with a hybrid approach of
prediction coding that exploits both spatial redundancy and
temporal redundancy of image signals. To exploit temporal
redundancy, motion estimation, and motion compensation theories
are introduced to video compression. The motion information is
encoded and transmitted along with residual information that
measures differences between predicted pixels and original pixels.
This approach is a foundation for various video coding standards
such as H.264/AVC \cite{wiegand2003overview}, H.265/HEVC \cite{sullivan2012overview}, and recent H.266/VVC
\cite{bross2021overview}. Following this hybrid approach, recent deep neural video
compression networks replace hand-crafted coding components
such as motion coding, and residual coding with corresponding
deep neural networks \cite{lu2019dvc, lu2020end, ding2021advances}. This motion coding and residual
coding-based deep neural video compression methods apply inter-prediction coding from traditional hybrid video codecs \cite{wiegand2003overview, sullivan2012overview, bross2021overview, han2021technical}. 
In detail, they use motion estimation and motion-compensated
prediction to generate the residual between the predicted frame and
the current encoding frame. Then, the residual, and the estimated
motion vector fields are encoded and transmitted to the decoder via
entropy coding. However, this approach has two main drawbacks.
Classical video coding applies motion coding and shows its
efficiency, thanks to its block-based approach. In other words,
motion coding shows its power only used together with a block-based approach. In conventional video coding, a frame is
partitioned into blocks via a partition structure such as a quad-tree
partition in HEVC, or a multi-type partition in VVC. The size of
the coding block varies from small block size such as 4$\times$4 to large
block size such as 64$\times$64. One block has only one motion vector for
P-frame coding and two motion vectors for B-frame coding. All
pixels inside the block share the same representative motion vector
of the block. In comparison with motion coding in deep neural
video compression, one motion vector of a block with a size of
16$\times$16 in classical video coding corresponds to 256 motion vectors
in deep neural video coding. Consequently, deep neural video
compressions require more bits to code motion information
compare to classical block-based hybrid video coding. In addition,
in classical block-based video coding such as HEVC, and VVC,
there are several efficient coding tools such as merge mode
estimation \cite{helle2012block, sullivan2012overview}, that can exploit motion redundancy
more deeply. For example, merge mode estimation can merge
neighboring blocks that share the same motion vector into a larger
region, and require a few bits to transmit only one representative
motion vector for the whole region and merge indices for sub-blocks in the region. This merge mode improves the coding
efficiency of HEVC from 6\% to 8\% as reported in \cite{helle2012block}. However,
deep neural video coding that works on pixel-level-based coding
does not exploit fully the capacity of motion coding as classical video coding can do. Consequently, the separation of motion 
coding and residual coding in deep neural video compression is not 
necessary to guarantee its high coding efficiency such as its 
counterparts in classical video coding. Vice-versa, even it can cause 
drawbacks for developing efficient deep neural video coding 
networks. In this paper, we propose a novel architecture that fully 
removes motion estimation and motion coding components in the 
network. In addition, it also does not use residual coding anymore. 
Instead, we consider the reconstructed frame as a synthesized frame 
from neighbor frames explicitly and from itself implicitly. In other 
words, on the encoder side, inputs contain two reconstructed frames 
of neighboring frames, one interpolated reference frame, and the 
current coding frame, concatenated together and feed directly into 
the encoder. Then the encoder transforms the inputs into latent 
variables that then are passed into entropy coding, and transmitted 
to the decoder side. On the decoder side, a convolutional kernel-based synthesizer generates the reconstructed frames from decoded 
latent variables and the reference frames. For each reconstructed 
pixel, the convolutional kernels-based synthesis method estimates 
three pairs of 1D convolution kernels and uses them to convolve 
with three corresponding reference frames including the 
interpolated reference frame generated from a video frame 
interpolation network and two reconstructed reference frames (the 
predecessor one and the successor one in display order). The pixel dependent 1D kernels capture both motion and intensity 
information required for the generation. Figure 1 shows differences 
between motion and residual coding P- frame neural video 
compression framework and the proposed kernel-based motion-free B-frame neural video compression. In our framework, we 
completely remove motion components in the network and use 
convolutional kernels to synthesize the reconstructed frames from 
latent variables transmitted from the encoder. The latent variables 
represent the current frame, the reference frames, and their 
correlations.
Auto-encoders (AEs) \cite{kingma2013auto} are powerful foundational generative 
modeling architectures for deep neural image, and video 
compression problems, however, the major drawback of AEs is that 
they often generate visual artifacts, particularly blurry in 
reconstructed samples \cite{bredell2023explicitly}. In this paper, the proposed 
convolutional kernels-based synthesizer on the decoder side is used 
to alleviate the blurry artifact in the reconstructed frames.
In summary, the primary contributions of the proposed methods are 
as follows:
\begin{itemize}
\item A  motion-free coding framework for B-frame coding of 
a deep neural video compression.
\item A kernel-based convolution at the decoder side of an 
auto-encoder to alleviate the blurry artifact that is a major 
drawback of AEs.
\item Experimental results demonstrate that the proposed
method outperforms the previous neural video 
compression models significantly in terms of coding 
efficiency.
\item The model size of the proposed network is much smaller than that of the previous motion coding-based models.
\end{itemize}

\begin{figure*}
\begin{center}
   \includegraphics[width=.7\linewidth]{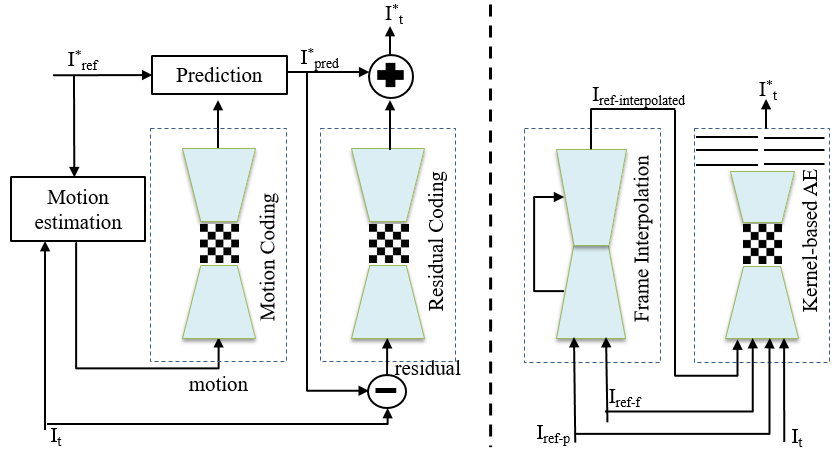}  
\end{center}
   \caption*{\hspace{10ex} (a) \hspace{42ex}  (b) }
   \caption{(a) Motion and residual coding-based neural video compression (b) The proposed kernel-based motion-free neural video compression.}
\label{fig:long}
\label{fig:onecol}
\end{figure*}

\section{Related Work}
\label{sec:formatting}

\subsection{Neural Image Compression}

Recently neural image compression \cite{iwai2024controlling, jiang2024neural} has attracted extensive 
research thanks to its impressive performance in comparison with 
traditional image codecs such as JPEG \cite{wallace1991jpeg}, BPG \cite{sullivan2012overview}, and VVC all-intra \cite{bross2021overview}. Start from the early work \cite{balle2016end} that proposed an end-to-end learned neural image compression model. Then \cite{theis2017lossy} introduced to use of compressive auto-encoders for lossy image compression, and \cite{balle2018variational} 
proposed to use of a VAE architecture with a scale hyper-prior network to 
boost coding efficiency. The VAE architecture becomes a baseline 
model for many following works in image compression \cite{minnen2018joint, he2021checkerboard, he2022elic, kim2022joint}. Therefore, in neural image compression, most works focus 
on improving entropy coding models via context models \cite{minnen2018joint, he2021checkerboard, he2022elic} and/or hierarchical hyper-prior models \cite{hu2020coarse, kim2022joint}, using 
convolutional networks or transformers. Recently transformer-based image compressions \cite{zhu2021transformer, qian2022entroformer, liu2023learned} obtained very impressive 
results in comparison with convolutional neural networks. \cite{zhu2021transformer} proposed to use a transformer for transform coding 
meanwhile \cite{qian2022entroformer} applied a transformer for 
entropy coding. A mix of transformer-CNN architectures for image 
compression is explored in \cite{liu2023learned}.

\subsection{Neural Video Compression}

The mainstream video codecs such as AVC/H.264 \cite{wiegand2003overview}, 
HEVC/H.265 \cite{sullivan2012overview}, and VVC/H.266 \cite{bross2021overview} are typically encoded with 
one of three coding configurations: all-intra configuration, low-delay configuration, and random access configuration. In all-intra coding configuration, each frame is encoded using a neural image compression model separately. The current SOTA neural image compression \cite{liu2023learned} 
outperforms VVC all-intra coding.
Neural video compression \cite{ge2024task, van2024mobilenvc} with low-delay coding configuration is researched intensively with outstanding performances in terms of 
coding efficiency \cite{agustsson2020scale, hu2020coarse, li2022hybrid}, the 
early work DVC \cite{lu2019dvc} and its enhanced version DVCPro \cite{lu2020end} tried to 
replace coding components in traditional hybrid video coding with 
corresponding neural networks such as motion estimation networks, 
motion compensation network, entropy coding network… Most 
recent works \cite{agustsson2020scale,liu2020learned, ding2021advances, hu2021fvc, hu2022coarse} follow this motion coding and residual coding-based approach focusing on updating the structures of 
the component networks. Hu \textit{et al.} \cite{hu2021fvc} introduced motion-based coding framework on the feature space, Hu \textit{et al.} \cite{hu2022coarse} applied coarse-to-fine coding with hyperprior guided mode prediction, and Agustsson \textit{et al.} \cite{agustsson2020scale} proposed scale-space flow instead of an explicit motion 
estimation network for end-to-end video compression. In addition, Hu \textit{et al.}\cite{hu2020improving} proposed 
resolution adaptive motion coding to improve coding efficiency.
Besides predictive coding, conditional 
coding methods were presented in \cite{li2021deep, li2022hybrid, li2024neural} to exploit temporal relationships 
between frames. Li \textit{et al.}\cite{li2021deep} proposed to use a deep contextual 
model based on a conditional coding approach that utilizes 
temporal correlation. Li \textit{et al.} \cite{li2022hybrid}  extend the deep contextual 
model work by introducing a hybrid spatial-temporal entropy 
modeling to improve coding efficiency and obtain impressive 
results. 
However, both the predictive coding framework and the conditional 
coding framework still remains motion estimation, and motion 
coding networks as key components in their coding pipeline.

\subsection{Neural Video Compression with Video Frame Interpolation in Random Access Configuration}

Video frame interpolation or frame rate up-conversion is usually 
considered as a companion of video compression to improve 
coding efficiency in B-frame coding or post-processing engine after 
the decoder. Recently deep learning-based video frame 
interpolation has also seen impressive advances thanks to the 
breakthrough of neural network architectures such as adaptive 
convolution \cite{niklaus2017video_v1}, adaptive separable convolution \cite{niklaus2017video_v2}, feature 
refinement network \cite{kong2022ifrnet}, and transformer \cite{lu2022video, shi2022video}. The early work \cite{wu2018video}
uses keyframes encoded via a neural image compression to 
interpolate intermediate frames by using a frame interpolation 
network. Pourrera \textit{et al.} \cite{pourreza2021extending} applied the super-slo-mo video frame interpolation network \cite{jiang2018super} to boost the coding performance of B 
frame coding. Alexandre \textit{et al.} \cite{alexandre2023hierarchical} proposed a hierarchical B-frame coding network based on two-layer conditional augmented normalization flows with the support of video frame interpolation and super-resolution networks.

\section{Method}

A video compression problem can be expressed as an implicit 
formulation that takes as input the original frame, and references 
reconstructed frames. The implicit formulation represents 
correlations and relationships between frames. Our goal is then to 
learn a map that represents the current coding frame via the 
previously reconstructed frames. In the following section, we describe how 
our method generates pixel values of the current frame from the 
reference frames.
\subsection{Kernel-based Motion Free Auto-encoder for Video Compression}

\begin{figure}[t]
\begin{center}
   \includegraphics[width=.6\linewidth]{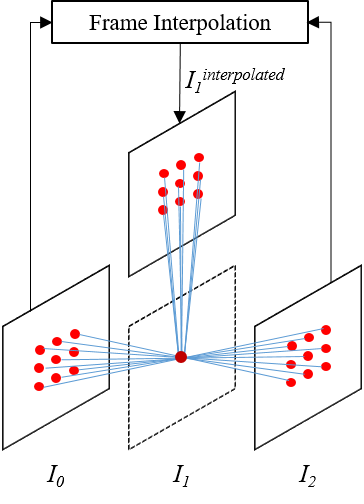}  
\end{center}
   \caption{Kernel-based pixel synthesis from reference pictures}
\end{figure}

In the paper, we propose a novel approach that removes motion 
estimation and motion coding by using a kernel-based auto-encoder. 
We do not estimate motion vectors and also do not separate motion 
coding and residual coding as in previous methods. We consider a 
reconstructed frame as a convolutional kernel-based synthesis from 
three reference frames, including the predecessor reconstructed 
frame, the successor reconstructed frame, and the interpolated 
frame from the two formers via a video frame interpolation 
network. Therefore our method contains only one auto-encoder for 
coding. The input of the network is a concatenation of the current 
encoding frame, denoted as \textit{I\textsubscript{1}} in Figure 2, and the three reference 
pictures, respectively marked as \textit{I\textsubscript{0}}, \textit{I\textsubscript{2}}, and \textit{I\textsubscript{1}\textsuperscript{interpolated}} in Figure 2. The high 
dimension of the stack of four frames is compressed into a latent 
variable space. The latent variables are coded and transmitted to the 
decoder. On the decoder side, inverse processes are applied to 
reconstruct the original frame. However, at the last layer of the 
decoder, we build three pairs of 1D kernels and use them to 
convolve with the corresponding reference frames to compute the value of the decoded pixels. For each reconstructed pixel \textit{I\textsubscript{1}(x,y)}, 
the network estimates three pairs of 1D kernels, 
denoted as 
\{\textit{k\textsubscript{0}\textsuperscript{v}(x,y)}, \textit{k\textsubscript{0}\textsuperscript{h}(x,y)}\},  
\{\textit{k\textsubscript{2}\textsuperscript{v}(x,y)}, \textit{k\textsubscript{2}\textsuperscript{h}(x,y)}\},
\{\textit{k\textsuperscript{v}\textsubscript{1-interpolated}(x,y)}, 
  \textit{k\textsuperscript{h}\textsubscript{1-interpolated}(x,y)}\}

where \textit{v} means vertical, and \textit{h} means horizontal then each pair of 
1D kernels is convoluted with a patch centered at \textit{(x,y)} in respective 
reference frames, denoted as \textit{P\textsubscript{0}(x,y)}, \textit{P\textsubscript{2}(x,y)}, \textit{P\textsubscript{1-interpolated}(x,y)}, the size 
of the patch is the same as that of the learned kernels. Then the final 
reconstructed pixel is the sum of outputs as the below equation:

\begin{equation}
 \begin{split}
 \textit{I\textsubscript{1}(x,y)} = \textit{k\textsubscript{0}\textsuperscript{v}(x,y)}*\textit{k\textsubscript{0}\textsuperscript{h}(x,y)}*\textit{P\textsubscript{0}(x,y)} +  \\ \textit{k\textsubscript{2}\textsuperscript{v}(x,y)}*\textit{k\textsubscript{2}\textsuperscript{h}(x,y)}*\textit{P\textsubscript{2}(x,y)} +  \\
    \textit{k\textsuperscript{v}\textsubscript{1-interpolated}(x,y)}*\textit{k\textsuperscript{h}\textsubscript{1-interpolated}(x,y)}*\textit{P\textsubscript{1-interpolated}(x,y)}    
\end{split}
\end{equation}  

This approach is novel in comparison with previous methods that 
often follow predictive coding and residual coding from classical 
hybrid video coding. Consider the decoded frame as the function
of the reconstructed neighbor frames explicitly and itself implicitly. In other words, the information of the current frame is encoded via auto-encoder-based image compression.

\begin{figure*}
\begin{center}
   \begin{subfigure}{.48\textwidth}
       \includegraphics[width=1\linewidth]{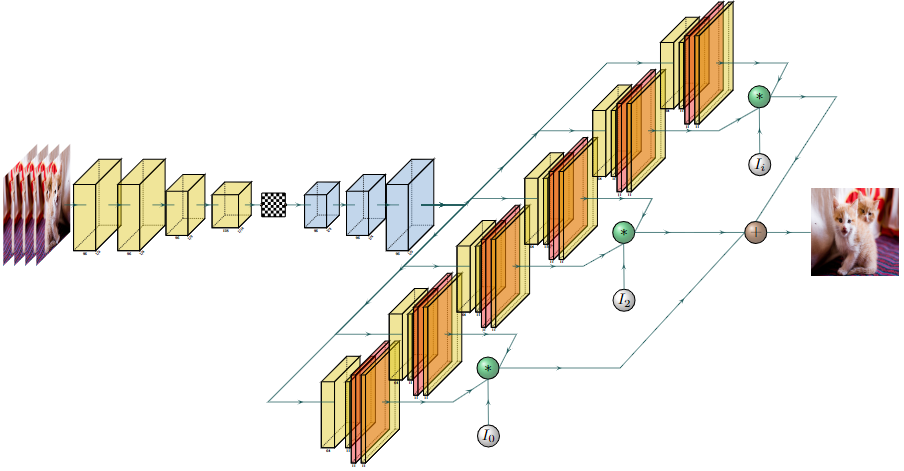}
       \caption{the proposed architecture}
   \end{subfigure}
   \begin{subfigure}{.48\textwidth}
       \includegraphics[width=1\linewidth]{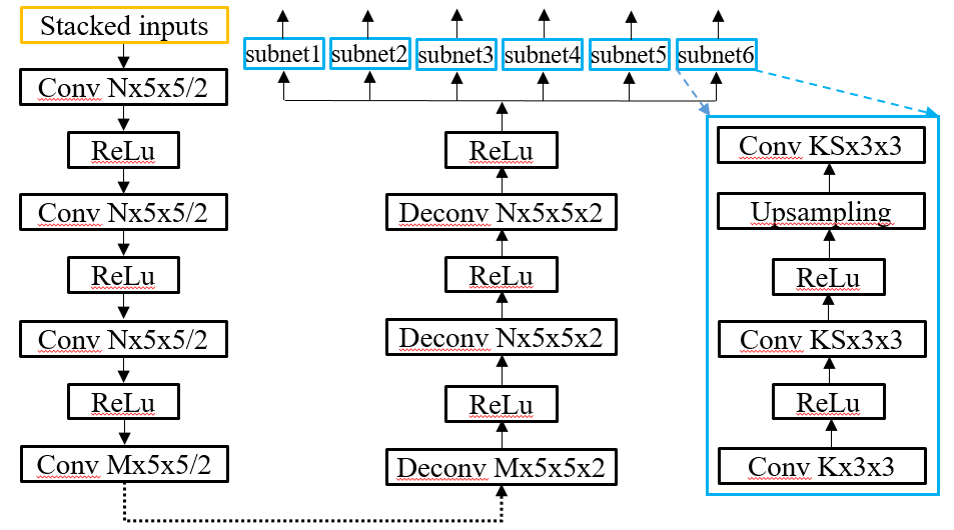}
       \caption{the detail layers of the network}
   \end{subfigure}
\end{center}
   \caption{ (a) Convolutional kernels-based motion-free AutoEncoder B-frame coding architecture. Six 1D convolutional kernels with 
kernel size of 31 in the last layer at the decoder side are convoluted with three reference frames, marked as I\textsubscript{0}, I\textsubscript{2}, I\textsubscript{i} (i stands for interpolated), two formers are the 
previously reconstructed frames, the other is the interpolated frame. (b) The detail layers of the proposed network with M = 128, N = 96, K = 64, and KS (kernel size) = 31, /2 means down-sampling stride of 2, and x2 means up-sampling stride of 2.}
\label{fig:long}
\label{fig:onecol}
\end{figure*}

\subsection{Hierarchical B-frame Coding with Video Frame Interpolation}
Among three reference frames used to synthesize the current 
coding frame at the decoder side, one is an interpolated frame 
generated from two other reference frames via a video frame 
interpolation network. In terms of temporal relation, compared to 
the other reference frames, the interpolated frame is the closest to
the current encoding frame. Therefore, it is very helpful for 
reconstructing the current encoding frame. However, due to its fake 
characteristics, it may have several visual artifacts. Consequently, it 
is necessary to apply both real reference frames and the generated 
frame that are complementary to each other. In other words, 
combining three reference frames for coding the current frame 
remains the effectiveness of both P-frame coding and B-frame 
coding.
Video frame interpolation problem is considered as a generative 
problem where an intermediate frame is generated from two 
neighboring frames, one is the predecessor frame and the other is 
the successor frame. As mentioned above, the video frame 
interpolation task has seen impressive results recently. Among 
many methods, we choose to use IFRnet \cite{kong2022ifrnet} with a small-size version thanks to its tradeoff between performance and lightweight model size. Note that we can apply any other video frame interpolation in our framework, just to generate an additional reference frame. In fact, we froze the interpolation network during the training of the video compression networks.
B-frame coding is a key component in classical video compression 
due to its full exploitation of temporal redundancy between frames 
in video sequences. In this paper, we combine B-frame coding with 
the frame interpolation network to improve coding efficiency thanks 
to recent impressive results in the video frame interpolation task.

Figure 4 shows an example of a coding structure of a Group of 
Pictures (GoP) with the size of 8. In this coding structure, I-frame 
coding \cite{balle2018variational} is used for coding the beginning frame in the display order of the GoP. Other 
frames (i.e. frame 1 to frame 7 in display order of the GoP) are 
encoded and decoded via B-frame coding by the proposed method. 
Firstly an interpolated reference frame is generated from two 
previously reconstructed neighboring frames, one is the 
predecessor frame and the other is the successor frame in display 
order as shown by arrows in Figure 4. Then, all three reference 
frames together with the current encoding frame are concatenated 
to input for the kernel-based auto-encoder shown in Figure 3. 
Finally, on the decoder side, the reconstructed frame of the current 
encoding frame is synthesized from the three reference frames via 
convolutional kernels as shown in Figure 3. It means that the frame 
interpolation network exists at both the encoder side and decoder side 
in our architecture.
\begin{figure}[t]
\hspace*{-20ex}
\begin{center}
        \includegraphics[width=.4\textwidth]{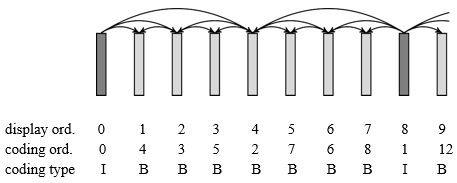}  
\end{center}
   \caption{A hierarchical B-frame Coding structure}
\end{figure}

\subsection{Loss Function and Entropy Coding}

Following the previous deep neural video compression methods \cite{agustsson2020scale, van2023hierarchical},
We apply a rate-distortion loss function that is a summation of two 
components: bitrate term and distortion term. The bitrate term 
represents the number of bits used to transmit information from 
encoder to the decoder, in our model, the bitrate term is estimated by 
using the well-known hyper-prior model \cite{balle2018variational} which is much 
simpler than context-based models proposed in recent works \cite{qian2022entroformer, liu2023learned}. The 
distortion term represents the difference between the reconstructed 
frame and the original frame, we use mean-square error to measure 
the difference. Besides two component terms, there is a scalar 
parameter applied to balance the trade-off between the two terms 
in the loss function, called lambda, denoted as $\lambda$.

\begin{equation}
    \textit{Loss} = \sum_{\substack{f  in  GoP}} (\lambda\textsubscript{\textit{f}}*D\textsubscript{\textit{f}}(I\textsubscript{\textit{f}}, \hat{I}\textsubscript{\textit{f}}) + R\textsubscript{\textit{f}}(\hat{Y}\textsubscript{\textit{f}}))
\end{equation}

in where, \textit{f} denotes the frame index, \textit{I\textsubscript{f}} is the current coding frame, $\hat{\textit{I}}\textsubscript{f}$ is the reconstructed frame, $D\textsubscript{\textit{f}}(I\textsubscript{\textit{f}}, \hat{I}\textsubscript{\textit{f}})$
is the distortion between \textit{I\textsubscript{f}}
and  is the quantized encoded feature,  is the bitrate 
estimated by the hyper-prior model \cite{balle2018variational}, and  is the trade-off 
parameter, lambda, for the corresponding frame \textit{f}. Basically, we 
can select independent lambda values for each frame. However, 
following traditional video coding that applies a hierarchical 
coding scheme for B-frame coding with Quantization Parameters 
(QP) levels, here we use hierarchical lambda values with three 
coding levels. In other words, for I-frames, we select a base lambda 
value, denoted as $\lambda\textsubscript{0}$ as level 0, for B-frames we derive lambda 
values $\lambda\textsubscript{1} = 0.85*\lambda\textsubscript{0}$ as level 1, and $\lambda\textsubscript{2} = 0.7*\lambda\textsubscript{0}$ as level 2.  For 
training, we use a GoP with a size of 5 frames with coding type I\textsubscript{0}B\textsubscript{2}B\textsubscript{1}B\textsubscript{2}I\textsubscript{0}.  Subscript numbers represent coding levels and 
receive corresponding lambda values during the training 
processing.

\section{Experimental Results}
\subsection{Dataset}
Following previous methods in deep neural video compression, we 
trained our networks on RGB images but evaluated at YUV420 
sequences that are often tested with classical video coding 
standards such as HEVC, and VVC. For training, we use the 
Vimeo90K dataset \cite{xue2019video} that contains nearly 90K tuples of seven 
sequential frames in RGB format that is suitable for various tasks 
of video processing, particularly for video compression. Three well-known UVG dataset \cite{mercat2020uvg}, HEVC – class B dataset \cite{bossen2013common}, and MCL-JCV dataset \cite{wang2016mcl}, that 
are in YUV420 format are used for evaluation. The UVG dataset 
contains seven video sequences at 120 fps. The 
HEVC Class B dataset includes five video sequences at various 
frame rates. The MCL-JCV dataset provides thirty video sequences at various frame rates. All test video sequences of three datasets are full-HD resolutions (1920x1080).
\subsection{Training}
Before training the video compression network, we train the IFRnet 
model for the video frame interpolation network with Vimeo90K 
dataset first. The video frame interpolation network is trained with 
a batch size of 8, during training, patches with a size of 224x224 
are cropped from training samples with a size of 448x256. Then 
we froze the IFRnet when training the whole video compression 
networks, including the I-frame coding network, and the B-frame 
coding network. We observed that a kernel size of 31 pixels performs 
well with a much faster training time when compared to larger kernel 
sizes. During training, we use a GoP with a size of 5 frames with a 
coding type IBBBI, set the learning rate to 10\textsuperscript{-4}
, and patches with a 
size of 256x256 are extracted randomly from the training samples 
with a size of 448x256. We trained models at four lambda values 
equal to \{0.5, 1, 3, 5\}$\ast$ 10\textsuperscript{-2}
. The video compression network is 
trained with a batch size of 8, it takes 3 days with a Nvidia A100 
GPU.
\subsection{Evaluation}
To verify the effectiveness of the proposed method in comparison 
with other approaches such as motion coding-based video compression networks and conditional coding-based models, we
compare our kernel-based motion-free deep neural video 
compression model with previous models that use motion 
estimation and motion coding, including FVC \cite{hu2021fvc} that applies motion-based coding framework on the feature space, and DCVC \cite{li2021deep} that learns context with motion estimation and motion compensation based on conditional coding. Both FVC and DCVC obtained high coding efficiency in low-delay coding configuration. For random access coding configuration, two SOTA networks, motion-based B-EPIC model \cite{pourreza2021extending}, and conditional augmented normalizing
flows based TLZMC model \cite{alexandre2023hierarchical} are compared.  We use the GoP with a size of 8 frames 
for evaluating our model, for other methods, we use results reported 
by authors in their papers. 

\textbf{Comparisons with the previous methods}. Figure 
5 shows the rate-distortion curves on the HEVC-class B dataset, 
the UVG dataset, and the MCL-JCV dataset respectively of various deep neural video 
compression models, including ours and the mentioned previous methods, the 
higher curve the better models’ performance is. As shown in 
Figure 5, for the HEVC-class B dataset our model outperforms the SOTA networks, but for the UVG dataset, TLZMC performs better than the others, including ours, for the MCL-JCV dataset, TLZMC does not report the result, our model is competitive with B-EPIC that requires high computations and DCVC that uses a specific 
complex context model for entropy coding that contributes 
significantly to its performance, meanwhile, our model applies the 
popular simple hyperprior model \cite{balle2018variational} reimplemented in 
compressAI library \cite{begaint2020compressai}. In addition, for MCL-JCV dataset at low bit-rate coding, our 
model is the best one and outperforms the previous motion-based 
models. It verifies our hierarchical coding scheme with frames at 
the highest level, they take less bitrate for coding because they are 
not used as reference frames of other frames.

\textbf{Bjøntegaard delta rate (BD-rate)} \cite{bjontegaard2001calculation} is a popular metric widely 
used in mainstream video coding standards to compare two video 
codecs, wherewith the same quality of the reconstructed frame, the 
BD-rate saving is measured, a lower bitrate means a better 
performance, therefore the relative BD-rate saving with a minus number represents a better codec. To verify the effectiveness of the kernel-based motion-free method, we carry 
out an experiment to compare the performance of the proposed 
kernel-based motion-free model with B-EPIC \cite{pourreza2021extending} that is the SOTA motion coding-based model with a symmetric auto-encoder in terms of BD-rate saving (\%) for PSNR with respect to the 
anchor SSF model \cite{agustsson2020scale}. Table 1 shows our kernel-based motion-free model provides a bitrate saving of 29.38\% in the 
UVG dataset, 32.75\% in the MCL-JCV, and the average 
bitrate saving is 31.07\% and outperforms B-EPIC.

\begin{table}
\begin{center}
\begin{tabular}{l|c|c|c}
\hline
Dataset          & UVG   &MCL-JCV  & Average \\
\hline
B-EPIC  & -28.5\%  & -24.21\%   &-26.36\%\\ 
ours    & \textbf{-29.38\%}  &\textbf{-32.75\%} &\textbf{-31.07\%}  \\ 

\hline
\end{tabular}
\end{center}
\caption{Comparisons of BD rate savings for RGB PSNR with respect to the anchor SSF model. Lower is better.}
\end{table}

\begin{figure*}
\centering
\begin{subfigure}{.33\textwidth}
  \includegraphics[width=1\linewidth]{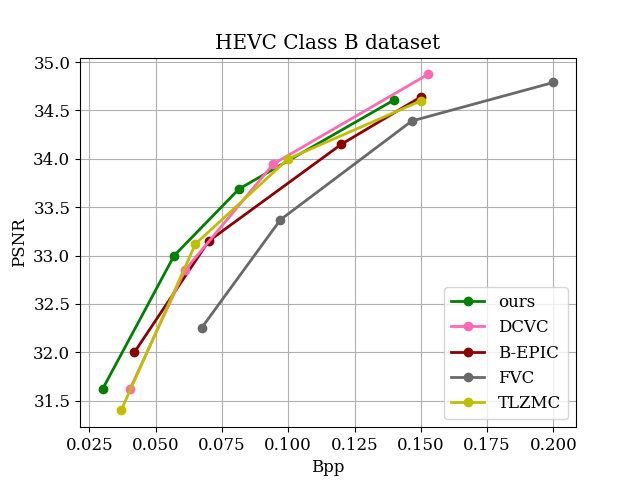}  
\end{subfigure} 
\begin{subfigure}{.33\textwidth}
  \includegraphics[width=1\linewidth]{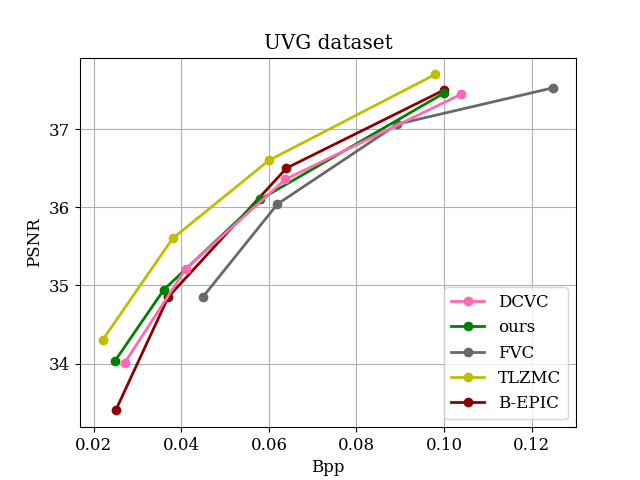}  
\end{subfigure}
\begin{subfigure}{.33\textwidth}
  \includegraphics[width=1\linewidth]{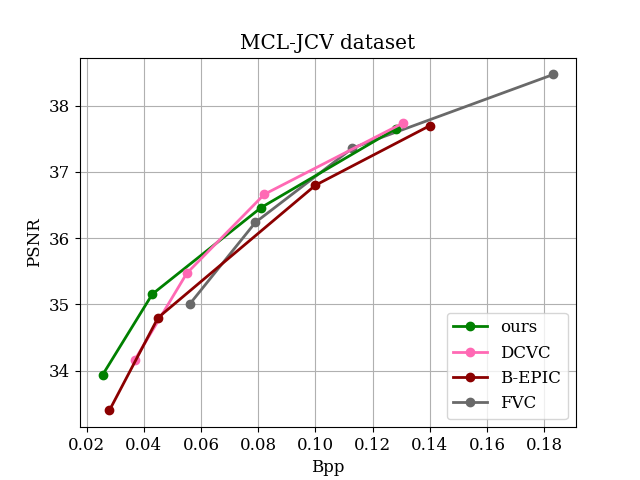}  
\end{subfigure}
\caption{Rate-distortion curves comparisons on the HEVC-class B, the UVG, and the MCL-JCV  datasets
}
\label{fig:2}
\end{figure*}

\subsection{Ablation Study}

To show the contribution of the third reference frame 
generated from a video frame interpolation network, we compare 
our final model with the model without a frame interpolation 
network. Consequently, the number of kernels is reduced from six 
to four due to the absence of the third reference frame. The performance of the final model increases 
significantly, with around 36.5\% bitrate saving compared to the anchor that is the model without frame interpolation network.

We carried out experiments with two kernel sizes 31 pixels, and 51 pixels. We found that when the kernel size is increased from 31 pixels to 51 pixels, the total parameters of six 1D kernel sub-networks increase from 0.49M to 0.65M but the rate-distortion curves do not change much. It means that the interpolated frame is quite "close" to the current encoding frame. In other words, pixels in a range of kernel size of 31 of the interpolated frame can estimate well the reconstructed pixel in the current encoding frame. 

\textbf{Alleviation of Blur artifacts compared to motion coding-based 
auto-encoder}: To illustrate the effectiveness of kernel-based 
coding on mitigating blurring artifacts, Figure 6 shows three 
examples from test sequences, the top row presents the example
frame and the cropped areas from the BQTerrace sequence of the HEVC-class B dataset, the next rows display the frames and the cropped areas
from the ReadySetGo sequence and the Jockey sequence of the UVG 
dataset. As shown in Figure 6, with a similar bitrate the 
conventional motion coding-based auto-encoder network with the representative model, SSF \cite{agustsson2020scale} usually generates 
visual artifacts, particularly blur in reconstructed frames. Our 
convolutional kernel-based motion-free auto-encoder network alleviates blur 
significantly thanks to six separated 1D convolutional kernels. For 
example, with the BQTerrace sequence that contains several periodic areas which are very 
challenging for reconstruction, the conventional motion coding-based auto-encoder often fails to handle those areas due to their 
ambiguous values and noisy characteristics. Meanwhile, our 
convolutional kernel-based motion-free auto-encoder alleviates 
significantly these ambiguous values and reconstructs pixels well 
in those challenging areas. This will open a new research direction for updating the foundational architecture of deep neural video 
compression.

\textbf{The size of component modules}: We present a component analysis in terms of model size in Table 3. The image codec to encode the beginning and ending frames of each GoP takes 38.04\% of the total size. Frame Interpolation network contributes 20.29\%, and six 1D kernel sub-networks only take 3.55\% of the total size. 

\subsection{Model Complexity and Running Time}

We compare the inference times of our model with the popular motion-based model, SSF \cite{agustsson2020scale}, both models are implemented by using the compressAI library \cite{begaint2020compressai}, running on the same environment on an Nvidia A100 GPU card. For SSF, we used the pretrained checkpoints provided by the library. As results shown in Table 2, our models run faster than the SSF models, around 1.7 times for full HD resolution video sequences. 

We compare the model complexity in model size and multiply-and-accumulate operations (MACs) with respect to two SOTA hierarchical random access coding models, B-EPIC \cite{pourreza2021extending}, and TLZMC \cite{alexandre2023hierarchical}. As shown in Table 4, the B-EPIC model has 54M parameters, the TLZMC model has 39.9M parameters meanwhile ours has only 13.8M, nearly four times smaller than B-EPIC and three times smaller than TLZMC, and a significant reduction in MACs compared to TLZMC. It verifies that our motion-free model can reduce the computational complexity significantly compared to motion-based counterparts.

\begin{table}
\begin{center}
\begin{tabular}{l|l|l|l}
\hline
Dataset        & resolution  & ssf (fps)   &ours (fps) \\
\hline
        HEVC-B & 1920x1080     &2.66  &\textbf{4.88}\\
        UVG    & 1920x1080     &2.96  &\textbf{4.93}\\
        MCL-JVC& 1920x1080     &2.76  &\textbf{4.5}         \\
\hline
        Average &1920x1080               &2.79  &\textbf{4.77} \\

\hline
\end{tabular}
\end{center}
\caption{Running time of our model vs that of the SSF model.}
\end{table}

\begin{table}[t]
\centering
\begin{tabular}{l|l|l}
    \hline
    Modules & Size & Ratio \\
    \hline
    Image codec & 5.25M & 38.04\% \\
    \hline
    Frame Interpolator & 2.8M & 20.29\%  \\
    \hline
    Six 1D kernel sub-networks & 0.49M & 3.55\% \\
    Frame auto-encoder &1.57M &11.38\% \\
    Frame Hyper-prior network & 3.69M & 26.74\% \\
    \hline
    Total & 13.8M  \\
    \hline

\end{tabular}
\caption{The size of component modules of the proposed model.}
\label{table2}
\end{table}

\begin{table}
\begin{center}
\begin{tabular}{l|l|l}
\hline
Model   & Size  &MACs \\
\hline
B-EPIC  & 54M  & - \\ 
TLZMC   & 39.9M &1.50 M/px \\ 
ours    & \textbf{13.8M}  & \textbf{0.65 M/px}\\ 

\hline
\end{tabular}
\end{center}
\caption{Computational complexity comparison with other random access coding models}
\end{table}

\begin{figure*}
\begin{center}
     \begin{subfigure}[b]{1\textwidth}
         \includegraphics[width=1\linewidth]{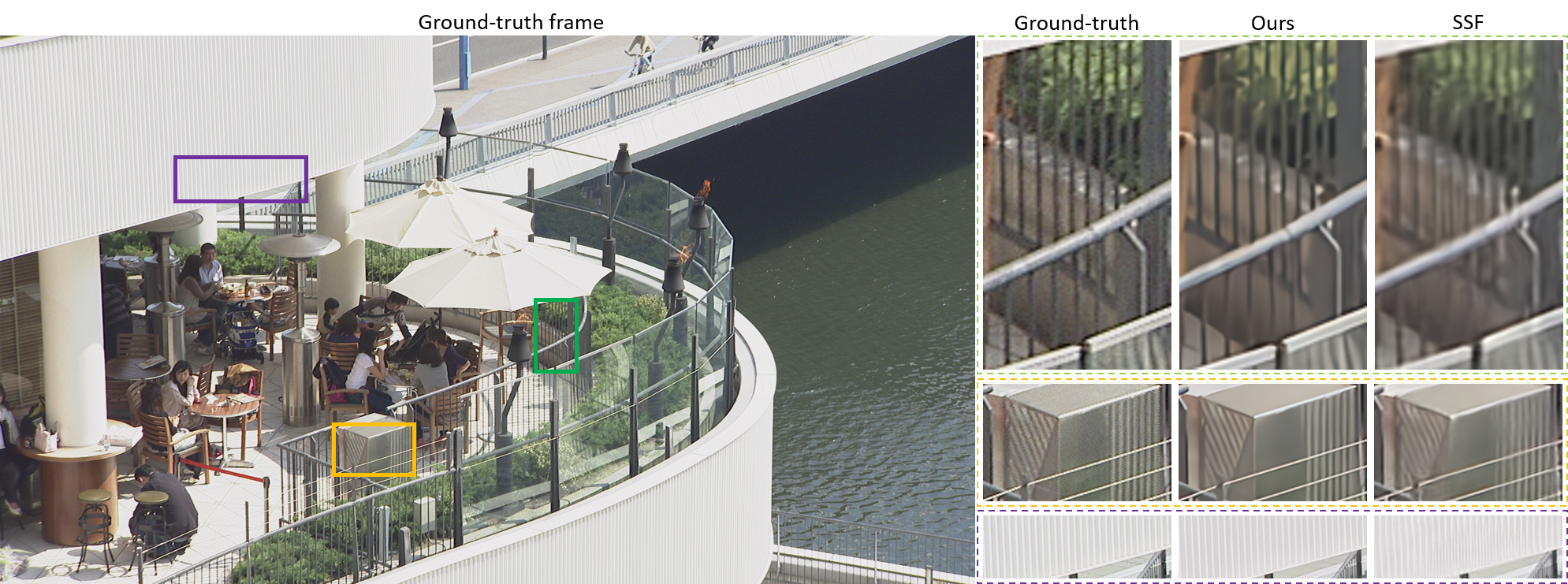}
     \end{subfigure}
     \begin{subfigure}[b]{0.497\textwidth}
         \includegraphics[width=1\linewidth]{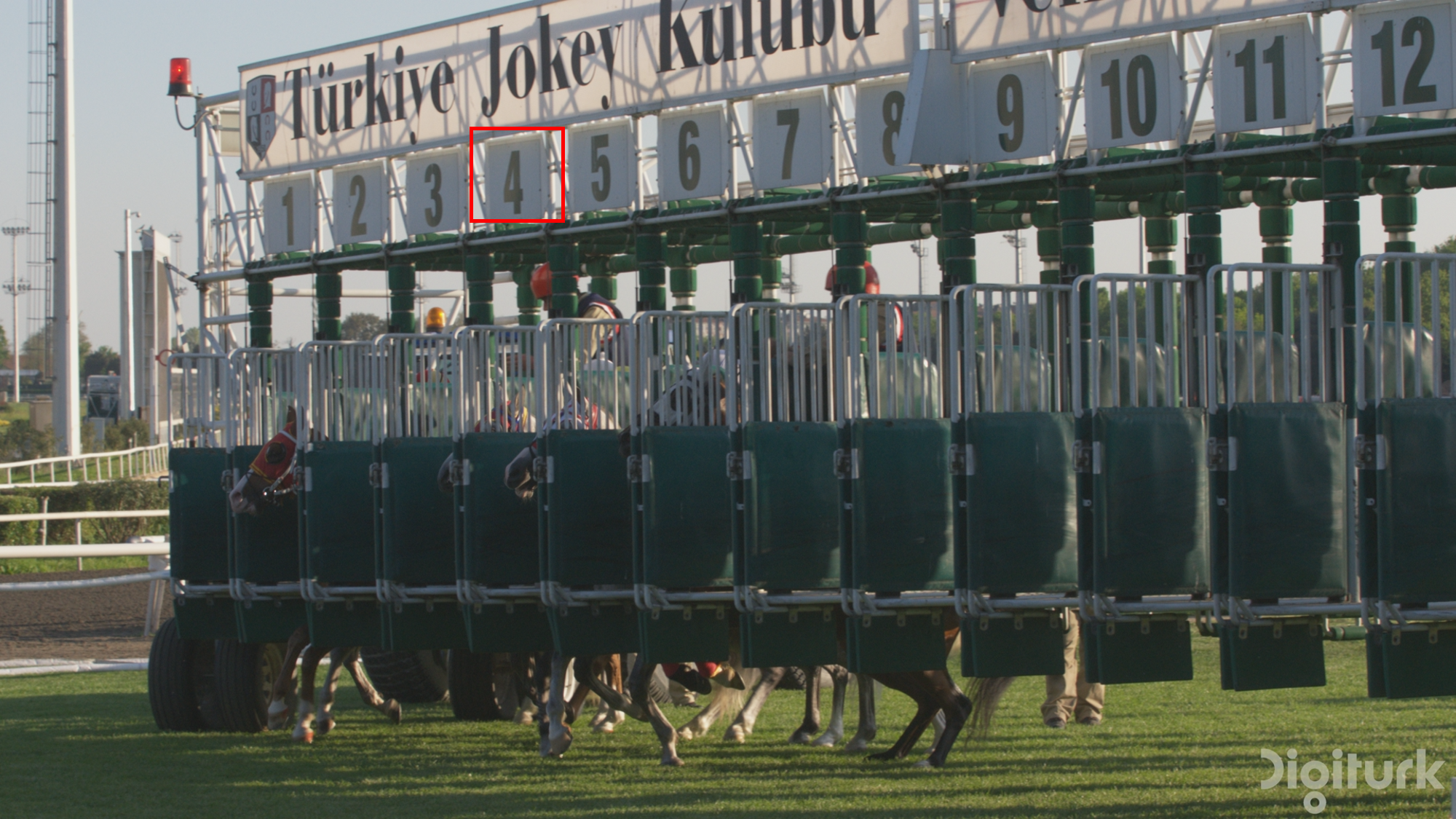}
     \end{subfigure} 
     \begin{subfigure}[b]{0.497\textwidth}
         \includegraphics[width=1\linewidth]{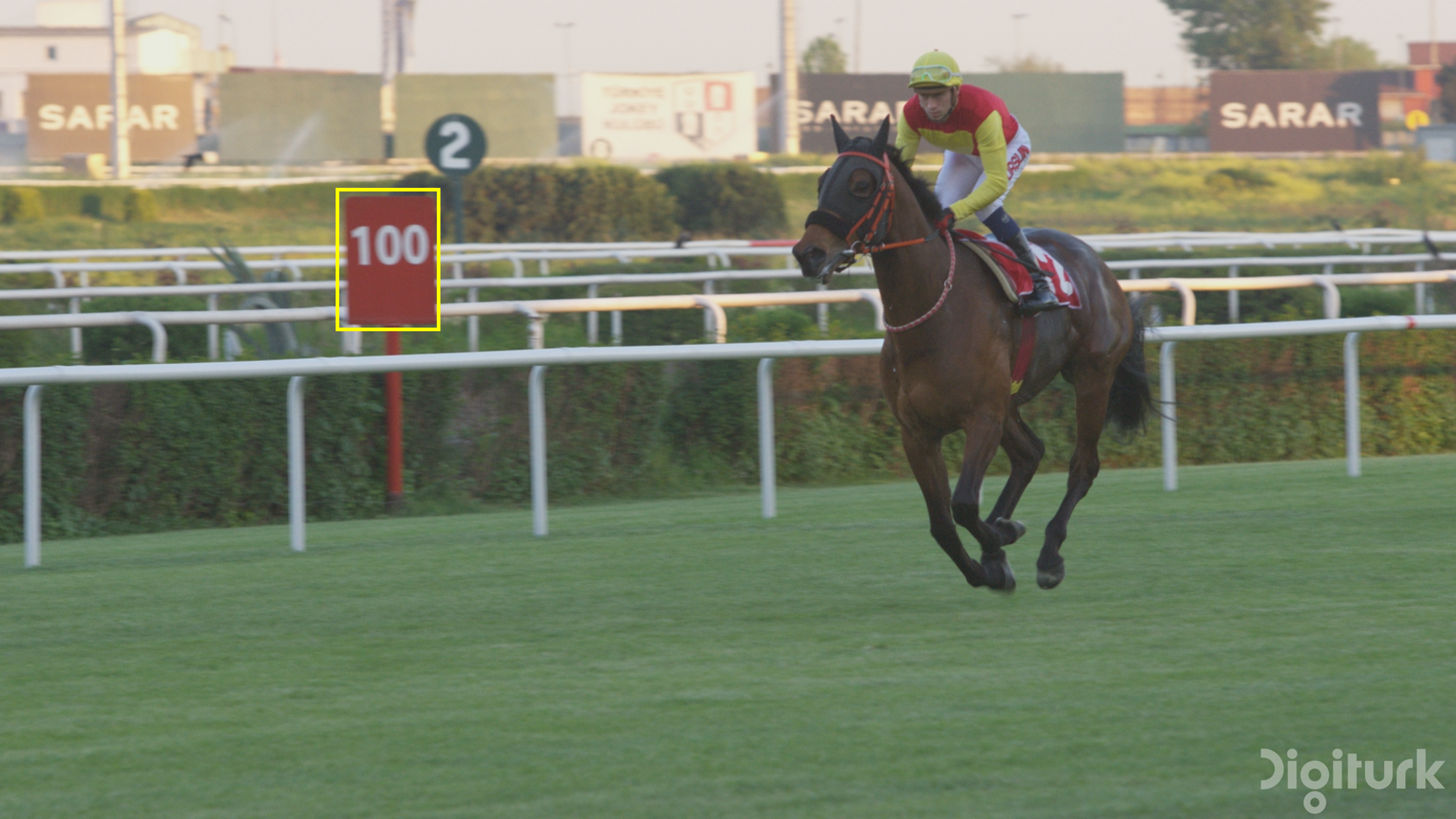}
     \end{subfigure}
          \begin{subfigure}[b]{0.497\textwidth}
         \includegraphics[width=1\linewidth]{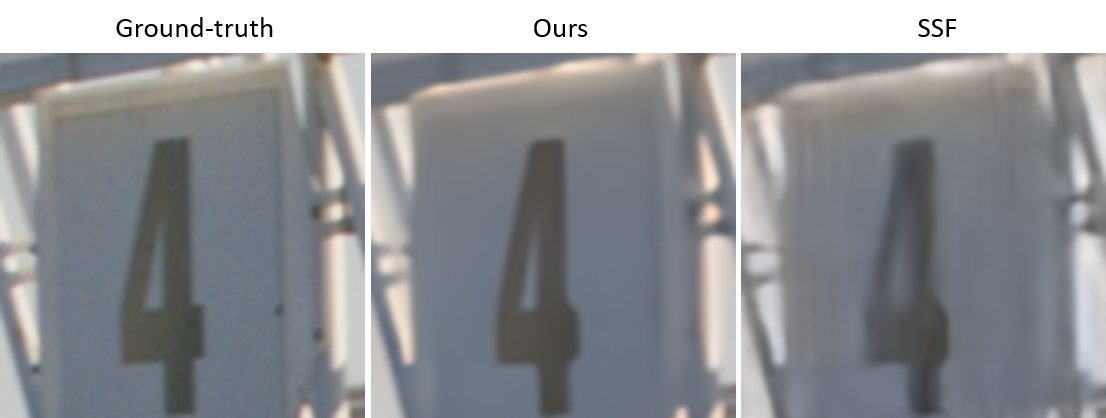}
     \end{subfigure} 
     \begin{subfigure}[b]{0.497\textwidth}
         \includegraphics[width=1\linewidth]{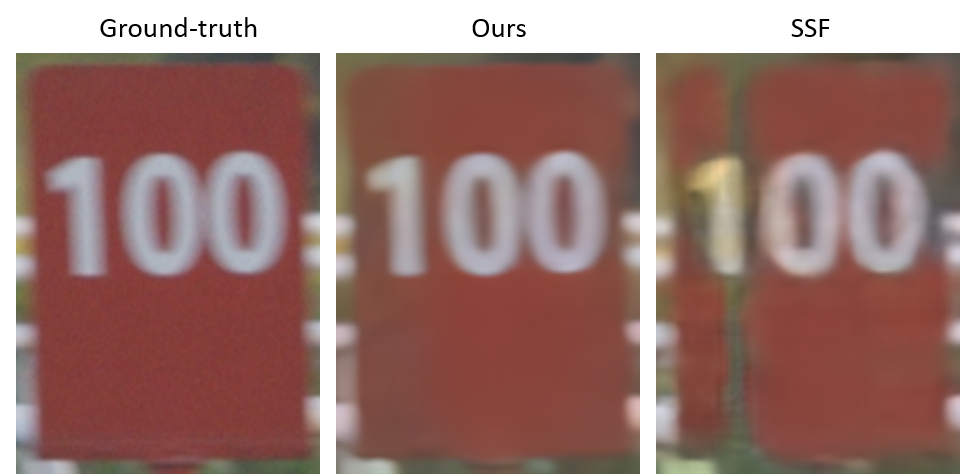}
     \end{subfigure}
\end{center}
     \caption{Visual comparisons with the motion-based SSF model on the HEVC-class B and the UVG datasets}
\end{figure*}
\section{Conclusion}

In this paper, we propose a novel kernel-based motion-free B-frame 
neural video coding that removes motion components in the coding. 
In addition, we introduce a convolutional kernel-based frame 
generator that synthesizes the reconstructed frame directly from 
reference frames via six 1D convolutional kernels. One of the 
reference frames is generated from the previously decoded 
reference frames using a video frame interpolation network. 
Our method outperforms the previous motion-coding-based deep 
neural video compressions meanwhile its model
size is much smaller than those of motion-based networks. It opens a new direction for future 
research that exploits temporal redundancy without motion 
estimation and motion coding. In addition our convolutional kernels-based 
autoencoder also alleviates blur artifacts that are often occurred in 
the conventional autoencoder. This encourages future research to 
explore new architectures of autoencoders for video compression.

{\small
\bibliographystyle{ieee_fullname}
\bibliography{egbib}
}

\end{document}